%% file: S-ebvpAoSbL.tex
\documentclass{article}

\usepackage{common}

\begin{document}

\title{\normalfont\bfseries\large State/event based versus purely Action or State based Logics}
\author{James Smith\\\texttt{james.smith@djalbat.com}}
\date{}
	
\maketitle

\begin{abstract}
\noindent Although less studied than purely action or state based logics, state/event based logics are becoming increasingly important. Some systems are best studied using structures with information on both states and transitions, and it is these structures over which state/event based logics are defined. The logic UCTL and its variants are perhaps the most widely studied and implemented of these logics to date. As yet, however, no-one seems to have defined UCTL*, a trivial step but a worthwhile one. Here we do just that, and define mappings that preserve truth between this logic and its more commonplace fragments CTL* and ACTL*. Also, acknowledging the importance of modal transition systems, we define a state/event based logic over a modified modal transition system as a precursor to further work.
\end{abstract}

\include*{introduction}

\include*{definitions2}

\include*{mappings2}

\include*{definitions3}

\include*{conclusions}

\pagebreak

\bibliographystyle{plain}
\bibliography{references}

\end{document}

%% file: introduction.tex
\section{Introduction}

We define the logic UCTL* over Kripke transition systems~\cite{DBLP:conf/sas/Muller-OlmSS99}, otherwise known as doubly labelled transition systems~\cite{DBLP:conf/lics/NicolaV90a}. As suggested in~\cite{DBLP:conf/litp/NicolaV90}, our Kripke transition systems and labelled transition systems~\cite{DBLP:conf/lics/LarsenT88} carry sets of actions rather than just a single action on each transition, making mappings between these structures and Kripke structures simpler. We define ACTL*~\cite{DBLP:conf/litp/NicolaV90} and ACTL\footnote{There is more than one logic given the name ACTL in the literature. The only one we refer to is the branching time logic for labelled transition systems proposed in~\cite{DBLP:conf/litp/NicolaV90}.} over these structures. We make small changes to the action formulae of~\cite{DBLP:conf/litp/NicolaV90} to support sets of actions, bringing them in line with those of~\cite{DBLP:journals/scp/BeekFGM11}. 

Inspired by~\cite{DBLP:conf/litp/NicolaV90}, we show some details of the proofs that the mappings between ACTL* and CTL* preserve truth and add mappings from UCTL* to both ACTL* and CTL*. In a similar vein we show some details of the proofs that the mappings between ACTL and CTL preserve truth and add mappings from UCTL~\cite{DBLP:journals/scp/BeekFGM11} to both ACTL and CTL. 

We also briefly look at 3-valued logics.  We define a variant of Kripke modal transition systems~\cite{DBLP:conf/esop/HuthJS01} which again carries sets of actions on each transition rather than just a single action. In order to accomodate this change, we replace $must$ and $may$ transitions~\cite{Schmidt} with $!$ and $?$ modifiers on the actions. We define the 3-valued logic UPML, a variant of 3-valued PML~\cite{DBLP:conf/cav/BrunsG99} which includes features of 3-valued PML\textsuperscript{\texttt{Act}}~\cite{DBLP:conf/vmcai/GodefroidJ03}.

%% file: definitions2.tex
\section{Definitions for 2-valued logics}

We define the structures for 2-valued logics, some common concepts, and the syntax and semantics and these logics. In the case of labelled transition systems and Kripke transition systems, note that in the definitions which follow we limit the number of transitions between any two states in any one direction to at most one. 

\begin{definition}A labelled transition system or LTS is a tuple $(S,Act,{\longrightarrow})$ where:
\label{def_lts}
\begin{itemize}
\item $S$ is a set of states ranged over by $s,s',s_0,s_1,...$,
\item $Act$ is set of actions ranged over by $a$ with $\alpha,\alpha_0,\alpha_1,...$ ranging over $2^{Act}$,
\item ${\longrightarrow}\subseteq S\times 2^{Act}\times S$ is the transition relation with $(s_0,\alpha,s_1)\in{\longrightarrow}$,
\item For any two transitions, $(s_0,\alpha_0,s_1),(s_0,\alpha_1,s_1)\in{\longrightarrow}\Rightarrow\alpha_0=\alpha_1$.
\end{itemize}
\end{definition}

\begin{definition}A Kripke structure or KS is a tuple $(S,{\longrightarrow},AP,\mathcal{L})$ where:
\label{def_ks}
\begin{itemize}
\item $S$ is a set of states ranged over by $s,s',s_0,s_1,...$,
\item ${\longrightarrow}\subseteq S\times S$ is the transition relation with $(s_0,s_1)\in{\longrightarrow}$,
\item $AP$ is a set of atomic propositions ranged over by $p$,
\item $\mathcal{L}:S\times AP{\longrightarrow}\{true,false\}$ is an interpretation function that associates a value of $true$ or $false$ with each $p\in AP$ for each $s\in S$.
\end{itemize}
\end{definition}

\begin{definition}A Kripke Transition System or KTS is a tuple $(S,Act,{\longrightarrow},AP,\mathcal{L})$ where:
\label{def_kts}
\begin{itemize}
\item $S$ is a set of states ranged over by $s,s',s_0,s_1,...$,
\item $Act$ is set of actions ranged over by $a$ with $\alpha,\alpha_0,\alpha_1,...$ ranging over $2^{Act}$,
\item ${\longrightarrow}\subseteq S\times 2^{Act}\times S$ is the transition relation with $(s_0,\alpha,s_1)\in{\longrightarrow}$,
\item $AP$ is a set of atomic propositions ranged over by $p$,
\item $\mathcal{L}:S\times AP{\longrightarrow}\{true,false\}$ is an interpretation function that associates a value of $true$ or $false$ with each $p\in AP$ for each $s\in S$,
\item For any two transitions, $(s_0,\alpha_0,s_1),(s_0,\alpha_1,s_1)\in{\longrightarrow}\Rightarrow\alpha_0=\alpha_1$.
\end{itemize}
\end{definition}

\noindent Note that since transitions carry sets of actions and not just one, there is no silent action $\tau$. Instead the empty set $\{\}$ is considered silent.

For convenience we overload $\mathcal{L}$ and define, for each $s\in S$, the function $\mathcal{L}(s):AP{\longrightarrow}\{true,false\}$ where $\mathcal{L}(s)(p)=\mathcal{L}(s,p)$. The set $\{\mathcal{L}(s)\;|\;s\in S\}$ is ranged over by $\omega,\omega_0,\omega_1,...$ and on occasion we write these functions in the form $\{p\mapsto true,...\}$ rather than $\{(p,true),...\}$. It is also useful to define, for some $\alpha\in 2^{Act}$ and $\omega\in\{\mathcal{L}(s)\;|\;s\in S\}$, the transformations $\alpha'=\{a\mapsto true\;|\;a\in\alpha\}\cup\{a\mapsto false\;|\;a\notin\alpha\}$ and $\omega'=\{p\;|\;p\mapsto true\in\omega\}$.

Paths are sequences of transitions where the final state of one transition equals the initial state of the next transition, if there is one. They are ranged over by $\sigma,\sigma'$ and $\sigma''$. For a KS, $\sigma=(s_0,s_1)(s_1,s_2)...$ whereas $\sigma=(s_0,\alpha_0,s_1)(s_1,\alpha_1,s_2)...$ for a KTS or LTS. Maximal paths are either infinite or their last state has no outgoing transitions. For the set of maximal paths starting at state $s$ we write $\mu path(s)$. For the initial and final states of the first transition of a path $\sigma$ we write $\,_\mathsf{S}(\sigma)$ and $(\sigma)_\mathsf{S}$, respectively. For a KTS or LTS, we write $(\sigma)_\mathsf{T}$ for the set of actions of the first transition of a path $\sigma$. Usually we abbreviate these with $\,_\mathsf{S}\sigma$, $\sigma_\mathsf{S}$ and $\sigma_\mathsf{T}$, respectively. A suffix $\sigma'$ of a path $\sigma$ is such that $\sigma=\sigma''\sigma'$ for some possibly zero length path $\sigma''$. A proper suffix $\sigma'$ of a path $\sigma$ is such that $\sigma=\sigma''\sigma'$ for some non-zero length path $\sigma''$. We write $\sigma\leqslant\sigma'$ when $\sigma'$ is a suffix of $\sigma$ and $\sigma<\sigma'$ when $\sigma'$ is a proper suffix of $\sigma$.

The part time logics ACTL\textsuperscript{-} and UCTL\textsuperscript{-} are introduced. Their path formulae will be redefined later in this section to form the logics ACTL and UCTL, respectively. In what follows $\phi$ and $\phi'$ are state formulae, $\pi$ and $\pi'$ are path formulae.

\begin{definition}The syntaxes of the logics CTL, CTL*, ACTL\textsuperscript{-}, ACTL*, UCTL\textsuperscript{-} and UCTL* are, with pleasing symmetry:

\[
\begin{array}{ll}
\begin{array}{ll}
\text{CTL}\\
&\begin{array}{rcl}
\;\;\;\,\phi&::=&p\;|\;\neg\phi\;|\;\phi\wedge\phi'\;|\;\exists\pi\\[4pt]
\;\;\;\,\pi&::=&\neg\pi\;|\;X\phi\;|\;\phi\; U\phi'\;|\;\phi\; W\phi'\\[8pt]
\end{array}
\end{array}
&
\begin{array}{ll}
\text{CTL*}\\
&\begin{array}{rcl}
\;\;\,\phi&::=&p\;|\;\neg\phi\;|\;\phi\wedge\phi'\;|\;\exists\pi\\[4pt]
\;\;\,\pi&::=&\phi\;|\;\neg\pi\;|\;\pi\wedge\pi'\;|\;X\pi\;|\;\pi\;U\pi'\\[8pt]
\end{array}
\end{array}
\\
\begin{array}{ll}
\text{ACTL\textsuperscript{-}}\\
&\begin{array}{rcl}
\phi&::=&true\;|\;\neg\phi\;|\;\phi\wedge\phi'\;|\;\exists\pi\\[4pt]
\pi&::=&\neg\pi\;|\;X\phi\;|\;X_a\phi\;|\;\phi\; U\phi'\;|\;\phi\;W\phi'\\[8pt]
\end{array}
\end{array}
&
\begin{array}{ll}
\text{ACTL*}\\
&\begin{array}{rcl}
\phi&::=&true\;|\;\neg\phi\;|\;\phi\wedge\phi'\;|\;\exists\pi\\[4pt]
\pi&::=&\phi\;|\;\neg\pi\;|\;\pi\wedge\pi'\;|\;X\pi\;|\;X_a\pi\;|\;\pi\; U\pi'\\[8pt]
\end{array}
\end{array}
\\
\begin{array}{ll}
\text{UCTL\textsuperscript{-}}\\
&\begin{array}{rcl}
\phi&::=&p\;|\;\neg\phi\;|\;\phi\wedge\phi'\;|\;\exists\pi\\[4pt]
\pi&::=&\neg\pi\;|\;X\phi\;|\;X_a\phi\;|\;\phi\;U\phi'\;|\;\phi\;W\phi'\\[8pt]
\end{array}
\end{array}
&
\begin{array}{ll}
\text{UCTL*}\\
&\begin{array}{rcl}
\phi&::=&p\;|\;\neg\phi\;|\;\phi\wedge\phi'\;|\;\exists\pi\\[4pt]
\pi&::=&\phi\;|\;\neg\pi\;|\;\pi\wedge\pi'\;|\;X\pi\;|\;X_a\pi\;|\;\pi\; U\pi'\\[8pt]
\end{array}
\end{array}
\end{array}
\]

\end{definition}

\noindent Note that for CTL, ACTL\textsuperscript{-}, and UCTL\textsuperscript{-}, only $\phi$ contributes to the formulae. For CTL*, ACTL* and UCTL*, both $\phi$ and $\pi$ contribute to the formulae.

\begin{definition} 

The semantics of CTL and CTL* are defined over Kripke structures, KS; ACTL and ACTL* over labelled transition systems, LTS; and UCTL and UCTL* over Kripke transition systems, KTS. Specifically:

\vspace{1em}

\noindent For CTL, CTL*, ACTL, ACTL*, UCTL and UCTL*:

\begin{itemize}
\item $s\models\neg\phi$ iff $s\nmodels\phi$,
\item $s\models\phi\wedge\phi'$ iff $s\models\phi$ and $s\models\phi'$,
\item $s\models\exists\pi$ iff $\exists\sigma\in\mu path(s):\sigma\models\pi$.
\end{itemize}

\noindent For CTL, CTL*, UCTL\textsuperscript{-} and UCTL*:

\begin{itemize}
\item $s\models p$ iff $\mathcal{L}(s)(p)=true$.
\end{itemize}

\noindent For CTL, CTL*, ACTL, ACTL*, UCTL and UCTL*:

\begin{itemize}
\item $\sigma\models\neg\pi$ iff $\sigma\nmodels\pi$.
\end{itemize}

\noindent For CTL, ACTL\textsuperscript{-} and UCTL\textsuperscript{-}:

\begin{itemize}
\item $\sigma\models X\phi$ iff $\sigma_\mathsf{S}\models\phi$.
\end{itemize}

\noindent For ACTL\textsuperscript{-} and UCTL\textsuperscript{-}:

\begin{itemize}
\item $\sigma\models X_a\phi$ iff $\sigma_\mathsf{S}\models\phi$ and $a\in\sigma_\mathsf{T}$.
\end{itemize}

\noindent For CTL*:

\begin{itemize}
\item $\sigma\models X\pi$ iff $\exists s,s',\sigma'':\sigma=(s,s')\sigma'',\sigma''\models\pi$.
\end{itemize}

\noindent For ACTL* and UCTL*:

\begin{itemize}
\item $\sigma\models X\pi$ iff $\exists(s,\alpha,s')\sigma'':\sigma=(s,\alpha,s')\sigma'',\sigma''\models\pi$.
\item $\sigma\models X_a\pi$ iff $\exists(s,\alpha,s')\sigma'':a\in\alpha,\sigma=(s,\alpha,s')\sigma'',\sigma''\models\pi$.
\end{itemize}

\noindent For CTL*, ACTL* and UCTL*:

\begin{itemize}
\item $\sigma\models\phi$ iff $\,_\mathsf{S}\sigma\models\phi$,
\item $\sigma\models\pi\wedge\pi'$ iff $\sigma\models\pi$ and $\sigma'\models\pi'$,
\item $\sigma\models\pi\;U\pi'$ iff $\exists\sigma'\geqslant\sigma:\sigma'\models\pi',\forall\sigma'':\sigma\leqslant\sigma''<\sigma':\sigma''\models\pi$.
\end{itemize}

\noindent For CTL, ACTL\textsuperscript{-} and UCTL\textsuperscript{-}:

\begin{itemize}
\item $\sigma\models\phi\;U\phi'$ iff $\exists\sigma'\geqslant\sigma:\,_\mathsf{S}\sigma'\models\phi',\forall\sigma'':\sigma\leqslant\sigma''<\sigma':\,_\mathsf{S}\sigma''\models\phi$
\item $\sigma\models\phi\;W\phi'$ iff $\sigma\models\phi\;U\phi'$ or $\forall\sigma''\geqslant\sigma:\,_\mathsf{S}\sigma''\models\phi$.
\end{itemize}

\end{definition}

\noindent The $\forall$ operator is defined in the usual fashion as $\forall\pi=\neg\exists\neg\pi$ where appropriate.

\begin{definition}Action formulae have the following syntax:
\[
\chi\::=\tau\;|\;a\;|\;\neg\chi\;|\;\chi\wedge\chi'
\]
\end{definition}
\begin{definition}Action formulae have the following semantics:
\begin{itemize}
\item $\alpha\models \tau$ iff $\alpha=\{\}$,
\item $\alpha\models a$ iff $a\in\alpha$,
\item $\alpha\models\neg\chi$ iff $\alpha\nmodels\chi$,
\item $\alpha\models\chi\wedge\chi'$ iff $\alpha\models\chi$ and $\alpha\models\chi'$.
\end{itemize}
\end{definition}

\noindent For ACTL* and UCTL* we derive the following operators:

\[
X_\chi\pi=\mathlarger\bigvee\{\bigwedge_{a\in\alpha}X_a\pi\;|\;\alpha\in 2^{Act},\alpha\models\chi\}
\]
\[
\begin{array}{rcl}
\pi\,_\chi U_{\chi'}\pi'&=&\left(\pi\wedge X_\chi true\right)U\left(\pi\wedge X_{\chi'}\pi'\right)\\[12pt]
\pi\,_\chi W_{\chi'}\pi'&=&\left(\pi \,_\chi U_{\chi'}\pi'\right)\vee\left(\pi \,_\chi Ufalse\right)\\[4pt]
\end{array}
\]

\noindent These definitions cannot be applied to ACTL and UCTL. Instead we define the syntax and semantics of the logics to include them. 

\begin{definition}The syntax of the path formulae for ACTL and UCTL is:

\[
\pi::=X_\chi\phi\;|\;\phi \,_\chi U_{\chi'}\phi'\;|\;\phi \,_\chi W_{\chi'}\phi'
\]

\end{definition}

\begin{definition}The semantics of the path formulae for ACTL and UCTL are:
\begin{itemize}
\item $\sigma\models X_\chi\phi$ iff $\sigma=\sigma'\sigma''$ with $\sigma'_\mathsf{T}\models\chi$ and $\sigma'_\mathsf{S}\models\phi$.
\item $\sigma\models\phi \,_\chi U_{\chi'}\phi'$ iff $\exists\sigma'>\sigma:\,_\mathsf{S}\sigma'\models\phi,\sigma'_\mathsf{T}\models\chi',\sigma'_\mathsf{S}\models\phi',\forall\sigma\leqslant\sigma''<\sigma':\,_\mathsf{S}\sigma''\models\phi,\sigma''_\mathsf{T}\models\chi$. 
\item $\sigma\models\phi \,_\chi W_{\chi'}\phi'$ iff $\sigma\models\phi \,_\chi U_{\chi'}\phi'$ or $\forall\sigma''\geqslant\sigma:\,_\mathsf{S}\sigma''\models\phi,\sigma''_\mathsf{T}\models\chi$.
\end{itemize}
\end{definition}

\noindent Note that the $W$ operator did not make it into the definition of ACTL in~\cite{DBLP:conf/litp/NicolaV90}. We introduce it here to bring the definition of ACTL into line with that of UCTL.

%% file: mappings2.tex
\section{Mappings for 2-valued logics}

We define mappings between ACTL* and CTL* in both directions and show some details of the proofs that they preserve truth. We then devise similar mappings from UCTL* to CTL* and ACTL*. We define mappings between ACTL and CTL in both directions and again show some details of the proofs that they preserve truth. And again we devise similar mappings from UCTL to CTL and ACTL. 

In what follows, $\mathsf{F}$ is a fresh atomic proposition. Where convenient, for some $\omega'$ we write $\mathcal{L}(s)'$ where $\omega=\mathcal{L}(s)$. For convenience we also define $\{\mathsf{F}\}'=\{\mathsf{F}\mapsto true\}\cup\{p\mapsto false\;|\;p\in AP\}$ and $\{\mathsf{F}\}'\cup\omega'=\{\mathsf{F}\mapsto true\}\cup\omega'$.

\begin{figure}[h]
\centering
\includegraphics[scale=0.66666666]{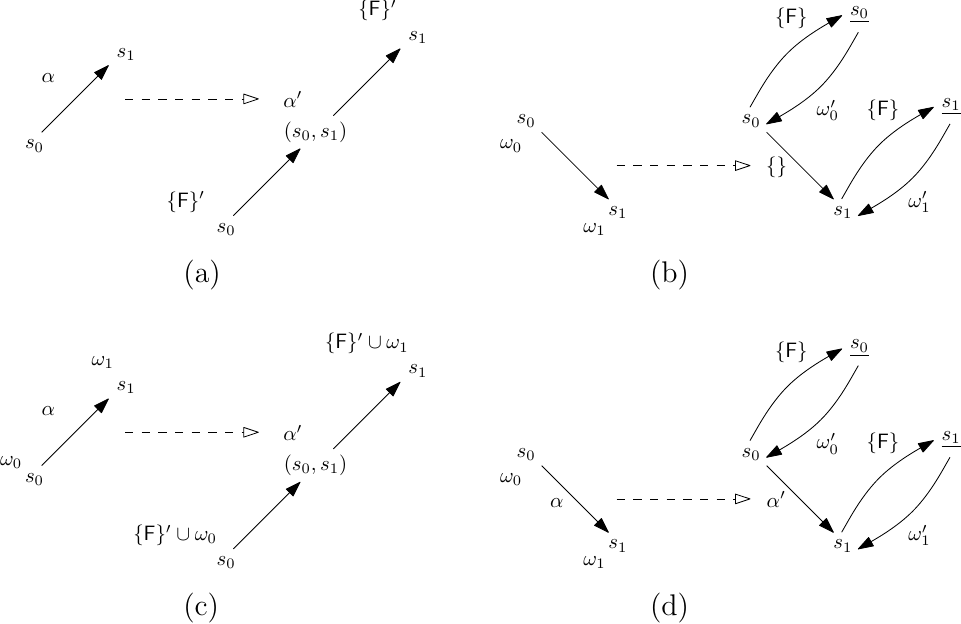}
\caption{Mappings for 2-valued logics}
\label{mappings}
\end{figure}

\vspace{-1.5em}

\subsection{$ks$, a mapping from ACTL* to CTL*}

Let $(S,Act,{\longrightarrow})$ be an LTS. The KS $(S',{\longrightarrow}',AP',\mathcal{L}')$ is defined:
\begin{itemize}
\item $S'=S\cup\{(s_0,s_1)\;|\;(s_0,s_1)\in{\longrightarrow}\}$,
\item $AP'=Act\cup\{\mathsf{F}\}$,
\item $\forall(s_0,\alpha,s_1)\in{\longrightarrow}:(s_0,(s_0,s_1))\in{\longrightarrow}'$ and $((s_0,s_1),s_1)\in{\longrightarrow}'$,
\item $\forall s\in S:\mathcal{L}'(s)=\{\mathsf{F}\}'$,
\item $\forall(s_0,\alpha,s_1)\in{\longrightarrow}:\mathcal{L}'((s_0,s_1))=\alpha'$.
\end{itemize}

\[
\begin{array}{cp{2em}c}
\begin{array}[t]{rcl}
ks(true)&=&true\\
ks(\neg\phi)&=&\neg ks(\phi)\\
ks(\phi\wedge\phi')&=&ks(\phi)\wedge ks(\phi')\\
ks(\exists\pi)&=&\exists ks(\pi)\\
\end{array}
&&
\begin{array}[t]{rcl}
ks(\neg\pi)&=&\neg ks(\pi)\\
ks(\pi\wedge\pi')&=&ks(\pi)\wedge ks(\pi')\\
ks(X\pi)&=&XX(ks(\pi))\\
ks(X_a\pi)&=&Xa\wedge XX(ks(\pi))\\
ks(\pi\;U\pi')&=&(\mathsf{F}\Rightarrow ks(\pi))U(\mathsf{F}\wedge ks(\pi'))
\end{array}
\end{array}
\]

\noindent Figure~\ref{mappings}(a) shows the construction.

\begin{theorem}
Let $L$ be an LTS with $\sigma$ a path in $L$ and $\phi$ an ACTL* formula, then the mapping $ks$ preserves truth, that is $L,\sigma\models\phi$ if and only if $ks(L),ks(\sigma)\models ks(\phi)$.
\label{ks_truth}
\begin{proof}
The proof is by induction on the length of the formula. Suppose $\phi,\phi',\pi,\pi'$ are all ACTL* formulae. We use the abbreviations $L_{ks}=ks(L)$, $\phi_{ks}=ks(\phi)$ and so on. The proof that $L,\sigma\models\pi\;U\pi'$ if and only if $L_{ks},\sigma_{ks}\models(\mathsf{F}\Rightarrow\pi_{ks})U(\mathsf{F}\wedge\pi'_{ks})$ is given. By definition, $\exists\sigma'\geqslant\sigma:L,\sigma'\models\pi',\forall\sigma\leqslant\sigma''\leqslant\sigma':L,\sigma''\models\pi$. By the induction hypothesis, $L,\sigma'\models\pi'$ if and only if $L_{ks},\sigma'_{ks}\models\pi'_{ks}$ and additionally, since $L_{ks},\sigma'_{ks}\models\mathsf{F}$ by construction $L,\sigma'\models\pi'$ if and only if $L_{ks},\sigma'_{ks}\models\mathsf{F}\wedge\pi'_{ks}$. Now consider $\sigma_{ks}\leqslant\sigma''_{ks}<\sigma_{ks}$. For those $\sigma''_{ks}$ with $\sigma''_{ks}=(\sigma'')_{ks}$ we have $L,\sigma''\models\pi$ if and only if $L_{ks},\sigma''_{ks}\models\pi_{ks}$ by the induction hypothesis and again by construction, $\sigma''_{ks}\models\mathsf{F}$, hence $L,\sigma''\models\pi$ if and only if $L_{ks},\sigma''_{ks}\models\mathsf{F}\Rightarrow\pi_{ks}$. For those $\sigma''_{ks}$ without, by construction $L_{ks},\sigma''_{ks}\nmodels\mathsf{F}$ and hence vacuously $L_{ks},\sigma''_{ks}\models\mathsf{F}\Rightarrow\pi_{ks}$. Hence $\forall\sigma\leqslant\sigma''<\sigma':L,\sigma''$ if and only if $\forall\sigma_{ks}\leqslant\sigma''_{ks}<\sigma'_{ks}:L_{ks},\sigma''_{ks}\models\mathsf{F}\Rightarrow\pi_{ks}$, which completes the given part of the proof.
\end{proof}
\end{theorem}

\subsection{$lts$, a mapping from CTL* to ACTL*}

Let $(S,{\longrightarrow},AP,\mathcal{L})$ be a KS. The LTS $(S',Act',{\longrightarrow}')$ is defined:
\begin{itemize}
\item $S=S\cup\{\underline{s}|s\in S\}$,
\item $Act'=AP\cup\{\mathsf{F}\}$,
\item ${\longrightarrow}'=\{(s_0,\{\},s_1)\;|\;(s_0,s_1)\in{\longrightarrow}\}\cup\{(s,\{\mathsf{F}\},\underline{s})\;|\;s\in S\}\cup\{(\underline{s},\{\mathcal{L}(s)'\},s)\;|\;s\in S\}$%
\end{itemize}

\[
\begin{array}{cp{2em}c}
\begin{array}[t]{rcl}
lts(p)&=&\exists(X_\mathsf{F}X_p true)\\
lts(\neg\phi)&=&\neg lts(\phi)\\
lts(\phi\wedge\phi')&=&lts(\phi)\wedge lts(\phi')\\
lts(\exists\pi)&=&\exists lts(\pi)
\end{array}
&&
\begin{array}[t]{rcl}
lts(\neg\pi)&=&\neg lts(\pi)\\
lts(\pi\wedge\pi')&=&lts(\pi)\wedge lts(\pi')\\
lts(X\pi)&=&X(\exists X_\mathsf{F}true\wedge lts(\pi))\\
lts(\pi\;U\pi')&=&(\exists X_\mathsf{F}true\wedge lts(\pi))U (\exists X_\mathsf{F}true\wedge lts(\pi'))
\end{array}
\end{array}
\]

\noindent Figure~\ref{mappings}(b) shows the construction.

\begin{theorem} 
Let $K$ be a KS with $\sigma$ a path in $K$ and $\phi$ a CTL* formula, then the mapping $lts$ preserves truth, that is $K,\sigma\models\phi$ if and only if $lts(K),lts(\sigma)\models lts(\phi)$.
\label{lts_truth}
\begin{proof}
The proof is by induction on the length of the formula. Let $K$ be a KS, then $K_{lts}=lts(K)$. The proof that $K,s\models p$ if and only if $K_{lts},s\models\exists(X_\mathsf{F} X_p true)$ is given. Suppose $K,s\models p$. By construction there are transitions $(s,\{\mathsf{F}\},s')$ and $(s',\omega',s)$ in $K_{lts}$ with $p\in\omega'$, therefore $K_{lts},s\models\exists(X_\mathsf{F}X_p true)$. Conversely, suppose $K_{lts},s\models\exists(X_\mathsf{F}X_p true)$. Then there must be transitions $(s,\omega'_1,s_1)$ and $(s_1,\omega'_2,s_2)$ with $\mathsf{F}\in\omega'_1$ and $p\in\omega'_2$. By construction $\mathsf{F}\in\omega'_1$ implies both $\omega'_1=\{\mathsf{F}\}$ and $s_1=s'$, however. Similarly by construction $s_2=s$ and since $\mathsf{F}\notin Act$, it can only be that this part of $K_{lts}$ corresponds to the state $s\in K$ with $s\models p$.
\end{proof}
\end{theorem}

\subsection{$ks_2$, a mapping from UCTL* to CTL*}

Let $(S,Act,{\longrightarrow},AP,\mathcal{L})$ be a KTS. The KS $(S',{\longrightarrow}',AP',\mathcal{L}')$ is defined:
\begin{itemize}
\item $S'=S\cup\{(s_0,s_1)\;|\;(s_0,s_1)\in{\longrightarrow}\}$,
\item $AP'=AP\cup Act\cup\{\mathsf{F}\}$,
\item $\forall(s_0,\alpha,s_1)\in{\longrightarrow}:(s_0,(s_0,s_1))\in{\longrightarrow}'$ and $((s_0,s_1),s_1)\in{\longrightarrow}'$,
\item $\forall s\in S:\mathcal{L}'(s)=\{\mathsf{F}\}\cup\mathcal{L}(s)'$,
\item $\forall(s_0,\alpha,s_1)\in{\longrightarrow}:\mathcal{L}'((s_0,s_1))=\alpha'$.
\end{itemize}

\[
\begin{array}{cp{2em}c}
\begin{array}[t]{rcl}
ks_2(p)&=&p\\
ks_2(\neg\phi)&=&\neg ks_2(\phi)\\
ks_2(\phi\wedge\phi')&=&ks_2(\phi)\wedge ks_2(\phi')\\
ks_2(\exists\pi)&=&\exists ks_2(\pi)\\
\end{array}
&&
\begin{array}[t]{rcl}
ks_2(\neg\pi)&=&\neg ks_2(\pi)\\
ks_2(\pi\wedge\pi')&=&ks_2(\pi)\wedge ks_2(\pi')\\
ks_2(X\pi)&=&XX(ks_2(\pi))\\
ks_2(X_a\pi)&=&Xa\wedge XX(ks_2(\pi))\\
ks_2(\pi\;U\pi')&=&(\mathsf{F}\Rightarrow ks_2(\pi))U(\mathsf{F}\wedge ks_2(\pi'))
\end{array}
\end{array}
\]

\noindent Figure~\ref{mappings}(c) shows the construction.

\begin{theorem}
Let $K$ be a KTS with $\sigma$ a path in $K$ and $\phi$ a UCTL* formula, then the mapping $ks_2$ preserves truth, that is $K,\sigma\models\phi$ if and only if $ks_2(K),ks_2(\sigma)\models ks_2(\phi)$.
\qed
\end{theorem}

\subsection{$lts_2$, a mapping from UCTL* to ACTL*}

Let $(S,Act,{\longrightarrow},AP,\mathcal{L})$ be a KTS. The LTS $(S',Act',{\longrightarrow}')$ is defined:
\begin{itemize}
\item $S=S\cup\{\underline{s}|s\in S\}$,
\item $Act'=Act\cup AP\cup\{\mathsf{F}\}$,
\item ${\longrightarrow}'=\{(s_0,\alpha,s_1)\;|\;(s_0,\alpha,s_1)\in{\longrightarrow}\}\cup\{(s,\{\mathsf{F}\},\underline{s})\;|\;s\in S\}\cup\{(\underline{s},\{\mathcal{L}(s)'\},s)\;|\;s\in S\}$%
\end{itemize}

\[
\begin{array}{cp{2em}c}
\begin{array}[t]{rcl}
lts_2(p)&=&\exists(X_\mathsf{F}X_p true)\\
lts_2(\neg\phi)&=&\neg lts_2(\phi)\\
lts_2(\phi\wedge\phi')&=&lts_2(\phi)\wedge lts_2(\phi')\\
lts_2(\exists\pi)&=&\exists lts_2(\pi)\\
\end{array}
&&
\begin{array}[t]{rcl}
lts_2(\neg\pi)&=&\neg lts_2(\pi)\\
lts_2(\pi\wedge\pi')&=&lts_2(\pi)\wedge lts_2(\pi')\\
lts_2(X\pi)&=&X(\exists X_\mathsf{F}true\wedge lts_2(\pi))\\
lts_2(X_a\pi)&=&X_a(lts_2(\pi))\\
lts_2(\pi\;U\pi')&=&(\exists X_\mathsf{F}true\wedge lts_2(\pi))U(\exists X_\mathsf{F}true\wedge lts_2(\pi'))
\end{array}
\end{array}
\]

\noindent Figure~\ref{mappings}(d) shows the construction.

\begin{theorem}
Let $K$ be a KTS with $\sigma$ a path in $K$ and $\phi$ a UCTL* formula, then the mapping $lts_2$ preserves truth, that is $K,\sigma\models\phi$ if and only if $lts_2(K),lts_2(\sigma)\models lts_2(\phi)$.
\qed
\end{theorem}

\subsection{$ks'$, a mapping from ACTL to CTL}

The mapping of structures is identical to the $ks$ mapping.

\[
\begin{array}{rcl}
ks'(true)&=&true\\
ks'(\neg\phi)&=&\neg ks'(\phi)\\
ks'(\phi\wedge\phi')&=&ks'(\phi)\wedge ks(\phi')\\
ks'(\exists\pi)&=&\exists ks'(\pi)\\
ks'(\neg\pi)&=&\neg ks'(\pi)\\
ks'(X_\chi\phi)&=&X(\neg\mathsf{F}\wedge\chi\wedge\exists X(\mathsf{F}\wedge ks'(\phi)))\\
ks'(\phi\,_\chi U_{\chi'}\phi')&=&((\mathsf{F}\wedge ks'(\phi))\vee(\neg\mathsf{F}\wedge\chi))U(\neg\mathsf{F}\wedge\exists((\neg\mathsf{F}\wedge\chi')U(\mathsf{F}\wedge ks'(\phi'))))\\
ks'(\phi\,_\chi W_{\chi'}\phi')&=&((\mathsf{F}\wedge ks'(\phi))\vee(\neg\mathsf{F}\wedge\chi))W(\neg\mathsf{F}\wedge\exists((\neg\mathsf{F}\wedge\chi')U(\mathsf{F}\wedge ks'(\phi'))))
\end{array}
\]

\begin{theorem}
Let $L$ be an LTS with $\sigma$ a path in $L$ and $\phi$ an ACTL formula, then the mapping $ks'$ preserves truth, that is $L,\sigma\models\phi$ if and only if $ks'(L),ks'(\sigma)\models ks'(\phi)$.
\label{ksprimed_truth}
\begin{proof}
The proof is by induction on the length of the formula. Let $L$ be an LTS with $\phi,\phi',\pi,\pi'$ formulae satisfied in $L$. Then $L_{ks'}=ks'(L)$, $\phi_{ks'}=ks'(\phi)$ and so on. The proof that $L,\sigma\models\phi_\chi U_{\chi'}\phi'$ if and only if $L_{ks'},\sigma_{ks'}\models((\mathsf{F}\wedge\phi_{ks'})\vee(\neg\mathsf{F}\wedge\chi))U(\neg\mathsf{F}\wedge\exists((\neg\mathsf{F}\wedge\chi')U(\mathsf{F}\wedge\phi'_{ks'})))$ is given. By definition, $L,\sigma\models\phi_\chi U_{\chi'}\phi'$ if and only if $\exists\sigma'>\sigma:L,\,_\mathsf{S}\sigma'\models\phi,L,\sigma'_\mathsf{T}\models\chi',L,\sigma'_\mathsf{S}\models\phi',\forall\sigma\leqslant\sigma''<\sigma':L,\,_\mathsf{S}\sigma''\models\phi,L,\sigma''_\mathsf{T}\models\chi$. Consider $\,_\mathsf{S}\sigma'$. By the induction hypothesis $L,\,_\mathsf{S}\sigma'\models\phi$ if and only if $L_{ks'},\,_\mathsf{S}(\sigma'_{ks'})\models\phi_{ks'}$ and by construction $\,_\mathsf{S}(\sigma'_{ks'})\models\mathsf{F}$, therefore $L,\,_\mathsf{S}\sigma'\models\phi$ if and only if $L_{ks'},\,_\mathsf{S}(\sigma'_{ks'})\models\mathsf{F}\wedge\phi_{ks'}$. Now consider $\sigma'_\mathsf{T}$ and $\sigma'_\mathsf{S}$. By construction $L,\sigma'_\mathsf{T}\models\chi'$ if and only if $L_{ks'},(\sigma'_{ks'})_\mathsf{S}\models\chi'$. Also, by the induction hypothesis $L,\sigma'_\mathsf{S}\models\phi'$ if and only if $L_{ks'},(\sigma'_\mathsf{S})_{ks'}\models\phi'_{ks'}$ and by construction $L_{ks'},(\sigma'_\mathsf{S})_{ks'}\models\mathsf{F}$ therefore $L,\sigma'_\mathsf{S}\models\phi'$ if and only if $L_{ks'},(\sigma'_\mathsf{S})_{ks'}\models\mathsf{F}\wedge\phi'_{ks'}$. Working in $L_{ks'}$, it remains to be shown that $(\sigma'_{ks'})_\mathsf{S}\models\chi'$ and $(\sigma'_\mathsf{S})_{ks'}\models\mathsf{F}\wedge\phi'_{ks'}$ if and only if $\,_\mathsf{S}((\sigma'_{ks'})_\mathsf{S},(\sigma'_\mathsf{S})_{ks'})\models\neg\mathsf{F}\wedge\exists((\neg\mathsf{F}\wedge\chi')U(\mathsf{F}\wedge\phi'_{ks'}))$. From left to right we have $(\sigma'_{ks'})_\mathsf{S}\models\neg\mathsf{F}$ hence $\,_\mathsf{S}((\sigma'_{ks'})_\mathsf{S},(\sigma'_\mathsf{S})_{ks'})\models\neg\mathsf{F}$. Similarly $(\sigma'_{ks'})_\mathsf{S}\models\neg\mathsf{F}\wedge\chi'$ and $(\sigma'_{ks'})_\mathsf{S}\models\mathsf{F}\wedge\phi'_{ks'}$ hence $\,_\mathsf{S}((\sigma'_{ks'})_\mathsf{S},(\sigma'_\mathsf{S})_{ks'})\models(\neg\mathsf{F}\wedge\chi')U(\mathsf{F}\wedge\phi'_{ks'})$. The result then follows. From right to left the argument is similar. Therefore $\sigma'_\mathsf{T}\models\chi'$ and $\sigma'_\mathsf{S}\models\phi$ if and only if $\,_\mathsf{S}((\sigma'_{ks'})_\mathsf{S},(\sigma'_\mathsf{S})_{ks'})\models\neg\mathsf{F}\wedge\exists((\neg\mathsf{F}\wedge\chi')U(\mathsf{F}\wedge\phi'_{ks'}))$. Lastly consider $\,_\mathsf{S}\sigma''$ and $\sigma''_\mathsf{T}$ with $\sigma_{ks'}\leqslant\sigma''_{ks'}<\sigma'_{ks'}$. For those $\sigma''_{ks'}$ with $\sigma''_{ks'}=(\sigma'')_{ks'}$ we have $L,\,_\mathsf{S}(\sigma'')\models\phi$ if and only if $L_{ks'},\,_\mathsf{S}(\sigma''_{ks'})\models\phi_{ks'}$ and by construction $L_{ks'},\,_\mathsf{S}(\sigma''_{ks'})\models\mathsf{F}$. Therefore $L,\,_\mathsf{S}\sigma''\models\phi$ if and only if $L_{ks'},\,_\mathsf{S}(\sigma''_{ks'})\models\mathsf{F}\wedge\phi_{ks'}$. For each of those $\sigma''_{ks'}$ without, by construction there is a unique $\sigma''$ with $L,\sigma''_\mathsf{T}\models\chi$ if and only if $L_{ks'},\,_\mathsf{S}(\sigma''_{ks'})\models\chi$. Also by construction $L_{ks'},\,_\mathsf{S}(\sigma''_{ks'})\nmodels\mathsf{F}$ and therefore $L,\sigma''_\mathsf{T}\models\chi$ if and only if $L_{ks'},\,_\mathsf{S}(\sigma''_{ks'})\models\neg\mathsf{F}\wedge\chi$. Hence $\forall\sigma\leqslant\sigma''<\sigma':\,_\mathsf{S}\sigma''\models\phi,\sigma''_\mathsf{T}\models\chi$ if and only if $\forall\sigma_{ks'}\leqslant\sigma''_{ks'}<\sigma'_{ks'}:\sigma''_{ks'}\models(\mathsf{F}\wedge\phi_{ks'})\vee(\neg\mathsf{F}\wedge\chi)$, which completes the given part of the proof.
\end{proof}
\end{theorem}

\subsection{$lts'$, a mapping from CTL to ACTL}

The mapping of structures is identical to the $lts$ mapping.

\[
\begin{array}{cp{2em}c}
\begin{array}[t]{rcl}
lts'(p)&=&\exists X_\mathsf{F}(\exists X_p true)\\
lts'(\neg\phi)&=&\neg lts'(\phi)\\
lts'(\phi\wedge\phi')&=&lts'(\phi)\wedge lts'(\phi')\\
lts'(\exists\pi)&=&\exists lts'(\pi)
\end{array}
&&
\begin{array}[t]{rcl}
lts'(\neg\pi)&=&\neg lts'(\pi)\\
lts'(X\phi)&=&X(lts'(\phi))\\
lts'(\phi\;U\phi')&=&(\exists X_\mathsf{F}true\wedge lts'(\phi))U(\exists X_\mathsf{F}true\wedge lts'(\phi'))\\
lts'(\phi\;W\phi')&=&(\exists X_\mathsf{F}true\wedge lts'(\phi))W(\exists X_\mathsf{F}true\wedge lts'(\phi'))
\end{array}
\end{array}
\]

\begin{theorem} 
Let $K$ be a KS with $\sigma$ a path in $K$ and $\phi$ a CTL formula, then the mapping $lts'$ preserves truth, that is $K,\sigma\models\phi$ if and only if $lts'(K),lts'(\sigma)\models lts'(\phi)$.
\label{ltsprimed_truth}
\qed
\end{theorem}

\subsection{$ks_2'$, a mapping from UCTL to CTL}

The mapping of structures is identical to the $ks_2$ mapping.

\[
\begin{array}[t]{rcl}
ks_2'(p)&=&p\\
ks_2'(\neg\phi)&=&\neg ks_2'(\phi)\\
ks_2'(\phi\wedge\phi')&=&ks_2'(\phi)\wedge ks_2(\phi')\\
ks_2'(\exists\pi)&=&\exists ks_2'(\pi)\\
ks_2'(\neg\pi)&=&\neg ks_2'(\pi)\\
ks_2'(X_\chi\phi)&=&X(\neg\mathsf{F}\wedge\chi\wedge\exists X(\mathsf{F}\wedge ks_2'(\phi'))\\
ks_2'(\phi\,_\chi U_{\chi'}\phi')&=&((\mathsf{F}\wedge ks_2'(\phi))\vee(\neg\mathsf{F}\wedge\chi))U(\neg\mathsf{F}\wedge\exists((\neg\mathsf{F}\wedge\chi')U(\mathsf{F}\wedge ks_2'(\phi'))))\\
ks_2'(\phi\,_\chi W_{\chi'}\phi')&=&((\mathsf{F}\wedge ks_2'(\phi))\vee(\neg\mathsf{F}\wedge\chi))W(\neg\mathsf{F}\wedge\exists((\neg\mathsf{F}\wedge\chi')U(\mathsf{F}\wedge ks_2'(\phi'))))
\end{array}
\]

\begin{theorem}
Let $K$ be a KTS with $\sigma$ a path in $K$ and $\phi$ a UCTL formula, then the mapping $ks'_2$ preserves truth, that is $K,\sigma\models\phi$ if and only if $ks'_2(K),ks'_2(\sigma)\models ks'_2(\phi)$.
\qed
\end{theorem}

\subsection{$lts_2'$, a mapping from UCTL to ACTL}

The mapping of structures is identical to the $lts_2'$ mapping.

\[
\begin{array}[t]{rcl}
lts_2'(p)&=&\exists X_\mathsf{F}(\exists X_p true)\\
lts_2'(\neg\phi)&=&\neg lts_2'(\phi)\\
lts_2'(\phi\wedge\phi')&=&lts_2'(\phi)\wedge lts_2'(\phi')\\
lts_2'(\exists\pi)&=&\exists lts_2'(\pi)\\
lts_2'(\neg\pi)&=&\neg lts_2'(\pi)\\
lts_2'(X_\chi\pi)&=&X_\chi(\exists X_\mathsf{F}true\wedge lts_2'(\pi))\\
lts_2'(\phi\,_\chi U_{\chi'}\phi')&=&(\exists X_\mathsf{F}true\wedge lts_2'(\phi))\,_\chi U_{\chi'}(\exists X_\mathsf{F}true\wedge lts_2'(\phi'))\\
lts_2'(\phi\,_\chi W_{\chi'}\phi')&=&(\exists X_\mathsf{F}true\wedge lts_2'(\phi))\,_\chi W_{\chi'}(\exists X_\mathsf{F}true\wedge lts_2'(\phi'))
\end{array}
\]

\begin{theorem}
Let $K$ be a KTS with $\sigma$ a path in $K$ and $\phi$ a UCTL formula, then the mapping $lts_2'$ preserves truth, that is $K,\sigma\models\phi$ if and only if $lts_2'(K),lts_2'(\sigma)\models lts_2'(\phi)$.
\qed
\end{theorem}

\begin{figure}[h]
\centering
\includegraphics[scale=0.75]{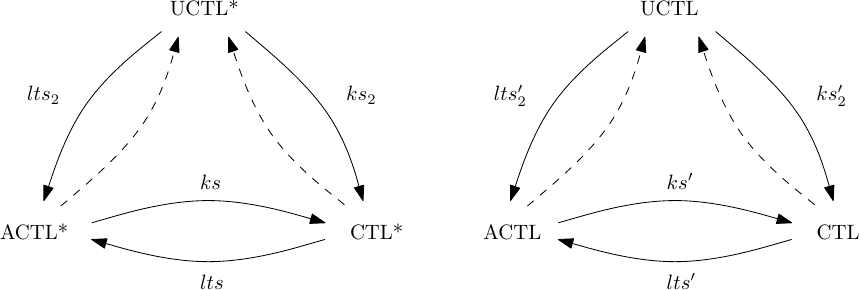}
\caption{A summary of the mappings for 2-valued logics}
\label{relationship2}
\end{figure}

%% file: definitions3.tex
`
\section{Definitions for 3-valued logics}

We define a variant of Kripke modal transition systems with $must$ and $may$ transitions replaced by modifiers on the actions, some common concepts, and then the syntax and semantics of the logic UPML. Note that in the definitions which follow we again limit the number of transitions between any two states in any one direction to at most one. 

\begin{definition}For a set of actions $Act$, a set of modified actions $Act_M$ is defined as $Act_M\subseteq Act\times\{!,?\}$. We write the elements of $Act_M$ in shorthand notation, for example $(a,!)$ becomes $a!$ and so on. For any set $Act_M$ we place restrictions on the elements, namely $a!\in Act_M\Rightarrow a?\notin Act_M$ or, equivalently, $a?\in Act_M\Rightarrow a!\notin Act_M$.
\end{definition}

\begin{definition}A Kripke modal transition system or KMTS is a tuple $(S,Act_M,{\longrightarrow},AP,\mathcal{L})$ where:
\label{def_ks}
\begin{itemize}
\item $S$ is a set of states ranged over by $s,s_0,s_1,...$,
\item $Act_M$ is set of modified actions ranged over by $a?,a!$ with $\alpha,\alpha_0,\alpha_1,...$ ranging over $2^{Act_M}$,
\item ${\longrightarrow}\subseteq S\times S$ is the transition relation with $(s_0,s_1)\in{\longrightarrow}$,
\item $AP$ is a set of atomic propositions ranged over by $p$,
\item $\mathcal{L}:S\times AP{\longrightarrow}\{true,\bot,false\}$ is an interpretation function that associates a value of $true$, $false$ or $\bot$, meaning unknown, with each $p\in AP$ for each $s\in S$,
\item For any two transitions, $(s_0,\alpha_0,s_1),(s_0,\alpha_1,s_1)\in{\longrightarrow}\Rightarrow\alpha_0=\alpha_1$.
\end{itemize}
\end{definition}

\noindent We overload $\mathcal{L}$ and define, for each $s\in S$, the function $\mathcal{L}(s):AP{\longrightarrow}\{true,\bot,false\}$ where $\mathcal{L}(s)(p)=\mathcal{L}(s,p)$. We also define, for $\alpha\in 2^{Act_M}$ and $\omega\in\{\mathcal{L}(s)\;|\;s\in S\}$, the transformations $\alpha'=\{a\mapsto true\;|\;a!\in\alpha\}\cup\{a\mapsto\bot\;|\;a?\in\alpha\}\cup\{a\mapsto false\;|\;a!\notin\alpha\wedge a?\notin\alpha\}$ and $\omega'=\{p!\;|\;p\mapsto true\in\omega\}\cup\{p?\;|\;p\mapsto\bot\in\omega\}$.

Paths and their related definitions are defined in an entirely analogous fashion to those for 2-valued structures.

To interpret propositional operators we use Kleene's strong 3-valued propositional logic~\cite{Kleene}. Negation maps $true$ to $false$, $false$ to $true$ and $\bot$ to $\bot$. The 3-valued truth table for conjunction and disjunction, the latter derived from the former by way of De Morgan's law $x\vee y=\neg(\neg x\wedge \neg y)$, is given below.

\[
\begin{array}{cccrrccccrrccccrr}
x&y&\;&x\wedge y&x\vee y&\quad\quad\quad&x&y&\;&x\wedge y&x\vee y&\quad\quad\quad&x&y&\;&x\wedge y&x\vee y\\[-8pt]
&&&&&&&&&&&&\\
\midrule\\[-8pt]
f&f&&f&f&&\bot&f&&f&\bot&&t&f&&f&t\\
f&\bot&&f&\bot&&\bot&\bot&&\bot&\bot&&t&\bot&&\bot&t\\
f&t&&f&t&&\bot&t&&\bot&t&&t&t&&t&t
\end{array}
\]

\noindent The logic UPML is introduced, which has the characteristics of both 3-valued PML~\cite{DBLP:conf/cav/BrunsG99} and 3-valued PML\textsuperscript{\texttt{Act}}.

\begin{definition}The syntax the logic UPML is:

\[
\phi::=p\;|\;\neg\phi\;|\;\phi\wedge\phi'\;|\;AX\phi\;|\;AX_a\phi
\]

\end{definition}

\begin{definition}The semantics of UPML are defined over Kripke modal transition sytems, KMTS. Specifically:

\[
\begin{array}{rcl}
\left[s\models\neg\phi\right]&=&\neg\left[s\models\phi\right]\\[4pt]
\left[s\models\phi\wedge\phi'\right]&=&\left[s\models\phi\right]\wedge\left[s\models\phi'\right]
\end{array}
\]
\[
[s\models AX\phi]=
\left\{
\begin{array}{ll}
true&\forall(s,\_,s'):[s'\models\phi]=true\\
false&\exists(s,\_,s'):[s'\models\phi]=false\\
\bot&otherwise
\end{array}
\right.
\]
\[
[s\models AX_a\phi]=
\left\{
\begin{array}{llrl}
true&\forall(s,\alpha,s'):&(a!\in\alpha\vee a?\in\alpha)\Rightarrow[s'\models\phi]&=true\\
false&\exists(s,\alpha,s'):&a!\in\alpha\wedge[s'\models\phi]&=false\\
\bot&otherwise
\end{array}
\right.
\]
\end{definition}

\noindent Here the underscore character $\_$ stands for any set of modified actions. The operators $EX$ and $EX_a$ can be derived in the usual manner. We give the semantics of the latter by way of an example, however:

\[
[s\models EX_a\phi]=
\left\{
\begin{array}{llrl}
true&\exists(s,\alpha,s'):&a!\in\alpha\wedge[s'\models\phi]&=true\\
false&\forall(s,\alpha,s'):&(a!\in\alpha\vee a?\in\alpha)\Rightarrow[s'\models\phi]&=false\\
\bot&otherwise
\end{array}
\right.
\]

\noindent Modifying actions with the $!$ and $?$ modifiers is equivalent to their transitions being $must$ and $may$ transitions, respectively. Note that this definition is a departure from convention~\cite{DBLP:conf/vmcai/GodefroidJ03,DBLP:conf/esop/HuthJS01} and that the correspondence is not quite straightforward, since all $must$ transitions are also $may$ transitions whereas actions are modified with only one of the $!$ and $?$ modifiers, not both. The correspondence is effectively a bijection, however, in terms of the above definitions. In the definitions of $[s\models AX_a\phi]$ and $[s\models EX_a\phi]$, for example, we see the term $(a!\in\alpha\vee a?\in\alpha)$ rather than just $a!\in\alpha$, making them entirely consistent with those of~\cite{DBLP:conf/esop/HuthJS01}.

Finally, we note that the $AX_a$ operator defined here has a different quality to the composite $\forall X_a$ operator defined in the 2-valued case. This has nothing to do with the 3-valued nature of the these logics nor the presence of $!$ and $?$ modifiers on the actions. Specifically, in the case of 2-valued logics, the $X_a\phi$ operator is satisfied by a path that is both labelled by an action $a$ \emph{and} who's next state satisfies $\phi$. Addition of the $\forall$ operator then ensures that all paths from a given state satisfy this condition. In the case of 3-valued logics, however, leaving aside the 3-valued nature of the logics and the modified actions, which do not affect the argument, the $AX_a$ operator is satisfied when all paths from a state satisfy the condition that labelling by an action $a$ \emph{implies} that the next state satisfies $\phi$.

%% file: conclusions.tex
\section{Conclusions}

We have defined the logic UCTL* over modified Kripke transition systems. Given recent developments~\cite{DBLP:journals/tosem/FantechiGLMPT12,DBLP:books/sp/sensoria2011/GnesiM11}, we claim this step is a worthwhile one. We have also modified Kripke modal transition systems in similar fashion and defined a logic, UPML, over these systems. We have defined mappings between the various logics in the 2-valued case. As reported in~\cite{DBLP:conf/lics/NicolaV90a}, the results of~\cite{DBLP:conf/litp/NicolaV90} led to practial gains in model checking at the time and it is interesting to note that, despite the plethora of model checkers around today, such mappings are still of practical use. For example, the mappings between UCTL, CTL and ACTL are used in the UMC model checker~\cite{DBLP:journals/scp/BeekFGM11}, which calls a minimisation program from the mCRL2 framework~\cite{DBLP:conf/dagstuhl/GrooteMRUW06} that works on labelled transition systems. The Kripke transition systems are therefore encoded as labelled transition systems as part of the minimisation process. Finally, we note that the results of~\cite{DBLP:conf/vmcai/GodefroidJ03} are not quite complete, their Kripke modal transition systems do not carry actions on their transitions, but a full investigation of the 3-valued case is left for future work.

\subsection{Acknowledgements}

My sincere thanks to Ian Hodkinson for his many helpful comments and corrections.